\documentclass [10pt] {article}

\usepackage[usenames]{color}

\usepackage [utf8] {inputenc}
\usepackage [T1] {fontenc}

\usepackage {epsfig}

\usepackage {url}
\usepackage {alltt}

\usepackage{amssymb}
\usepackage{amsmath}

\usepackage{listings}

\usepackage [pdftex,
bookmarksopen=true,
backref=false,
colorlinks=true,
linkcolor=blue,
urlcolor=blue,
citecolor=blue,
pdftitle={TechnicalReport},
pdfauthor={F. Boussinot},
pdfsubject={TechnicalReport}] {hyperref}

\newcommand {\email} [1] {{\href {mailto:#1} {#1}}}

\definecolor{MyDarkBlue}{rgb}{0,0,0.05}

{\vspace{0.3cm}
\scriptsize 
\lstset{
language=Java, 
numbers=right, 
numberstyle=\tiny, 
numbersep=-10pt, 
stepnumber=2, 
frame = single,
rulesepcolor=\color{MyDarkBlue}
}
#1}
{\vspace{0.3cm}}

\numberwithin {figure} {section}

\newcommand {\mvector} [1] {\overrightarrow {#1}}
\newcommand {\normal} [1] {norm ({#1})}

\newcommand {\degree} [1] {{#1}^{\circ}}

\begin {document}
\title { {\textbf {Determination of Forces from a Potential in
      Molecular Dynamics\\
(note)}}}

\newcommand {\fbsignature} {{\sc Fr\'ed\'eric Boussinot}\\
     Mines-ParisTech, Cemef\\
\small  {\email {frederic.boussinot@mines-paristech.fr}}}

\newcommand {\bmsignature} {{\sc Bernard Monasse}\\
    Mines-ParisTech, Cemef\\
\small  {\email {bernard.monasse@mines-paristech.fr}}
}

\author {\bmsignature \and \fbsignature}


\date {January 2014}

\maketitle
\begin {abstract}

  In Molecular Dynamics (MD), the forces applied to atoms derive from potentials which describe the energy of bonds, valence angles, torsion angles, and Lennard-Jones interactions of which molecules are made. These definitions are classic; on the contrary, their implementation in a MD system which respects the local equilibrium of mechanical conditions is usually not described. The precise derivation of the forces from the potential and the proof that their application preserves energy is the object of this note. This work is part of the building of a multi-scale MD system, presently under development.

\end {abstract}


\section {Introduction}

Numerical simulation at atomic scale predicts system states and properties from a limited number of physical principles, using a numerical resolution method implemented with computers. In Molecular Dynamics (MD)  \cite {MolecularDynamics} systems are organic molecules, metallic atoms, or ions. We concentrate on organic molecules, but our approach could as well apply to other kinds of systems. The goal is to determine the temporal evolution of the geometry and energy of atoms.

At the basis of MD is the classical (newtonian) physics, with the fundamental equation:

\begin {equation}
\mvector {F} = m \mvector {a} 
\end {equation}

where $\mvector {F}$ is the force applied to a particle of mass $m$ and $\mvector {a}$ is its acceleration (second derivative of the variation of the position, according to time).

A {\it force-field} is composed of several components, called {\it potentials} (of bonds, valence angles, dihedral angles, van der Waals contributions, electrostatic contributions, {\it etc}.) and  is defined by the analytical form of each of these components, and by the parameters caracterizing them. The basic components used to model molecules are the following: 

\begin {itemize}
\item atoms, with 6 degrees of freedom (position and velocity);

\item bonds, which link two atoms belonging to the same molecule; a bond between two atoms $a, b$  tends to maintain constant the distance $ab$.

\item valence angles, which are the angle formed by two adjacent bonds $ba$ et $bc$ in a same molecule; a valence angle tends to maintain constant the angle $\widehat{abc}$. A valence angle is thus concerned by the positions of three atoms.

\item torsion angles (also called {\it dihedral angles}) are defined by four atoms $a, b, c, d$ consecutively linked in the same molecule: $a$ is linked to $b$, $b$ to $c$, and $c$ to $d$; a torsion angle tends to priviledge particular angles between the planes $abc$ and $bcd$. These particular angles are the equilibrium positions of the torsion potential (minimal energies). In most cases, they are {\it Trans} (angle of $\degree {180}$), {\it Gauche} ($\degree {60}$) or {\it Gauche’} ($\degree {-60}$).

\item van der Waals interactions apply between two atoms which either belong to two different molecules, or are not linked by a chain of less than three (or sometimes, four) bonds, if they belong to the same molecule. They are pair potentials.

\end {itemize}

All these potentials depend on the nature of the concerned atoms and are parametrized differently in specific force-fields. Molecular models can also consider electrostatic interactions (Coulomb's law)  which are pair potentials, as van der Waals potentials are; their implementation is close to van der Waals potentials, with a different dependence to distance.

Intra-molecular forces (bonds, valence angles, torsion angles) as well as inter-molecular forces (van der Waals) are conservative: the work between two points does not depend on the path followed by the force between these two points. Thus, forces can be defined as derivatives of scalar fields. From now on, we consider that potentials are scalar fields and we have:

\begin {equation}
\mvector {F}(\mathbf r) = -\mvector {\nabla} {\cal U} (\mathbf r)
\end {equation}
where $\mathbf r$ denotes the coordinates of the point on which the force $\mvector {F}(\mathbf r)$ applies, and $\cal U$ is the potential from which the force derives.

The work presented here is part of a MD system presently under development\cite {MDRP}; the defined forces and their implementations have been tested with it.

\subsection* {Structure of the text}
Bonds are considered in Section \ref {section:liaison}, valence angles in Section \ref {section:valence}, torsion angles in Section \ref {section:torsion}, and finally Lennard-Jones potentials in Section \ref {section:lj}. Section \ref {section:resume} summarises the force definitions, and Section \ref {section:conclusion} concludes the text. A summary of the notations used in the paper is given in the Annex.

\section {Bonds \label {section:liaison}}
\begin{figure} [!htb]
\begin {center}
\includegraphics[width=10cm] {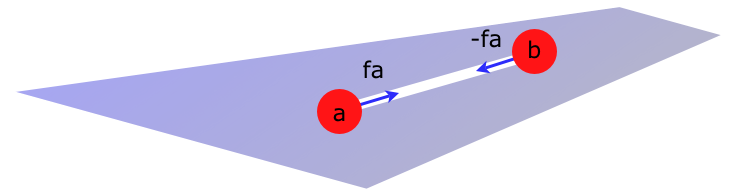}
\end {center}
\caption {Attractive bond between two linked atoms}
\label {figure:liaison}
\end{figure}

A bond models a sharing of electrons between two atoms which produces
a force between them. This force is the derivative of the bond
potential defined between the two atoms. Fig. \ref {figure:liaison}
shows a (attractive) force produced between two linked atoms $a$ and $b$. 

A {\it harmonic bond potential} is a scalar field $\cal
U$ which defines the potential energy of two atoms placed at distance
$r$ as:
\begin{equation}
{\cal U}(r) = k (r-r_0)^2
\end{equation}
where $k$ is the strength of the bond and $r_0$ is the equilibrium
distance (the distance at which the force between the two atoms is
null). We thus have:
\begin{equation}
{\frac {\partial {\cal U} (r)} {\partial {r}}} = 2k (r - r_0)
\end{equation}

\noindent
The partial derivative of $\cal U$ according to the position $r_a$ of
$a$ is:
\begin{equation}
{\frac {\partial {\cal U} (r)} {\partial {r_a}}} = {\frac {\partial
    {\cal U} (r)} {\partial r}} . {\frac {\partial {r}} {\partial {r_a}}} . 
\end{equation}

\noindent
But:
\begin{equation}
{\frac {\partial {r}} {\partial {r_a}}} = 1
\end{equation}

\noindent
We thus have:
\begin{equation}
{\frac {\partial {\cal U} (r)} {\partial r_a}}= 2k (r - r_0)
\end{equation}

Let $a$ and $b$ be two atoms, and $\mvector {u} = \normal {\mvector
  {ba}}$ be the normalization of vector $\mvector {ba}$. The force
produced on atom $a$ is:
\begin{equation} \label {eq:bond:fa} 
\mvector {f_a} = - {\frac {\partial {\cal U} (r)} {\partial r_a}} .\mvector {u} = -2k(r-r_0) .\mvector {u} 
\end{equation}
and the one on $b$ is the opposite, according to the action/reaction principle: 
\begin{equation} \label {eq:bond:fb} 
\mvector {f_b} = -\mvector {f_a}
\end{equation} 
Therefore, if $r > r_0$, the force on $a$ is a vector whose direction
is opposite to $\mvector {u}$ and tends to bring $a$ and $b$ closer
(attractive force), while it tends to bring them apart (repulsive
force) when $r < r_0$.

According to the definition of $\mvector {f_a}$ and $\mvector {f_b}$,
the sum of the forces applied to $a$ and $b$ is null (i.e. equilibrium
of forces):
\begin{equation} \label {eq:bond:null-sum} 
\mvector {f_a} + \mvector {f_b} = 0
\end{equation} 
Note that no torque is produced as the two forces are colinear.

\section {Valence Angles\label {section:valence}}

\begin{figure} [htb]
\begin {center}
\includegraphics[width=10cm] {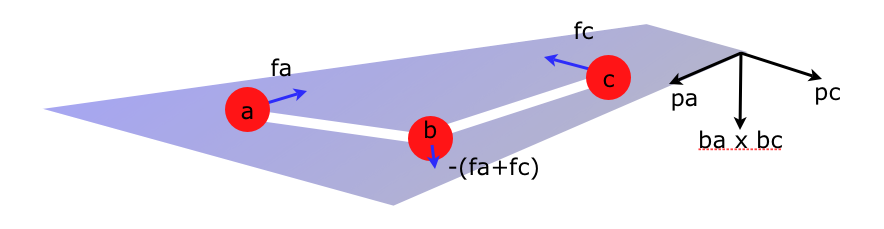}
\end {center}
\caption {Valence angle}
\label {figure:valence}
\end{figure}

Valence angles tend to maintain at a fixed value the angle between three atoms 
$a$, $b$ and $c$ such that $a$ is linked to $b$ and $b$ to $c$,
as shown on Fig. \ref  {figure:valence}.

The forces applied to the three atoms all belong to the plane $abc$
defined by the points $a$, $b$, $c$.

A {\it harmonic valence potential} is a scalar field $\cal U$ which
defines the potential energy of an atom configuration forming a
valence angle $\theta$ by:
 \begin{equation}
{\cal U}(\theta) = k (\theta-\theta_0)^2
\end{equation}
where $k$ is the strength of the valence angle and $\theta_0$ is the
equilibrium angle (the one for which energy is null). The partial
derivative of $\cal U$ according to the angle $\theta$ is thus:
\begin{equation}
{\frac {\partial {\cal U} (\theta)} {\partial \theta}}  = 2k (\theta-\theta_0)
\end{equation}

\noindent
The partial derivative of $\cal U$ according to the position $r_a$ of $a$ is:
\begin{equation}
{\frac {\partial {\cal U} (\theta)} {\partial r_a}}= 
{\frac {\partial {\cal U} (\theta)} {\partial \theta}} . 
{\frac {\partial \theta} {\partial r_a}}
\end{equation}
that is:
\begin{equation}
{\frac {\partial {\cal U} (\theta)} {\partial r_a}} = 2k
(\theta-\theta_0) . 
{\frac {\partial \theta} {\partial r_a}}
\end{equation}
As $a$ describes a circle with radius $|ab|$, centered on $b$, we
have\footnote {
The length of an arc of circle is equal to the product
  of the radius by the angle (in radians) corresponding to the arc of
  circle.
}:
\begin{equation}
{\frac {\partial \theta} {\partial r_a}} = \frac{1}{|ab|}
\end{equation}

\noindent
Let $\mvector {p_a}$ be the normalized vector in the plane $abc$,
orthogonal  to
$\mvector {ba}$~:
\begin{equation}
\mvector {p_a} = \normal {\mvector
    {ba} \times (\mvector {ba} \times \mvector {bc})}
\end{equation}

\noindent
The force applied on $a$ is then:
\begin{equation} \label {eq:valence:fa}
\mvector {f_a} = - {\frac {\partial {\cal U} (\theta)} {\partial
    {r_a}}} .\mvector {p_a} = -2k (\theta-\theta_0)/|ab| .\mvector {p_a}
\end{equation}
In the same way, the force applied on $c$ is:
\begin{equation} \label {eq:valence:fc}
\mvector {f_c} = -2k (\theta-\theta_0)/|bc|.\mvector {p_c}
\end{equation}
where $\mvector {p_c}$ is the normalized vector in plane $abc$,
orthogonal  to
$\mvector {cb}$~:
\begin{equation}
\mvector {p_c} = \normal {\mvector
    {cb} \times (\mvector {ba} \times \mvector {bc})}
\end{equation}

\noindent
The sum of the forces should be null:
\begin{equation}\label {eq:valence:null-sum}
\mvector {f_a} + \mvector {f_b} + \mvector {f_c}  = 0
\end{equation}
Thus, the force applied to $b$ is:
\begin{equation} \label {eq:valence:fb}
\mvector {f_b} = - \mvector {f_a} - \mvector {f_c}
\end{equation}

\subsection {Torques}\label {subsection:toques}
Let us now consider torques (moment of forces). The torque exerted by $\mvector {f_a}$ on
$b$ is $\mvector {ba} \times \mvector {f_a}$ and the torque exerted by $\mvector {f_c}$ on
$b$ is $\mvector {bc} \times \mvector {f_c}$. As $\mvector {ba}$ and
$\mvector {f_a}$ are orthogonal, one has\footnote{
If $u \bot v$ then $|u \times v| = |u||v|$.
}:

\begin{equation}
|\mvector {ba} \times \mvector {f_a}| = |\mvector {ba}| |\mvector
{f_a}| = |ba|  |-2k (\theta-\theta_0)/|ab|| = |2k (\theta-\theta_0)|
\end{equation}
For the same reasons:
\begin{equation}
|\mvector {bc} \times \mvector {f_c}| = |2k (\theta-\theta_0)|
\end{equation}
Thus, the two vectors $\mvector {ba} \times \mvector {f_a}$ and
$\mvector {bc} \times \mvector {f_c}$ have the same length. As they are by
construction in opposite directions, the
sum of the two torques on $b$ is null:
\begin{equation}\label{eq:valence:torque}
\mvector {ba} \times \mvector {f_a} + \mvector {bc} \times \mvector
{f_c} = 0
\end{equation}

\noindent
As a consequence, no rotation around $b$ can result from the
application of the two forces $\mvector {f_a} $ and $\mvector {f_c}$.

\section {Torsion Angles\label {section:torsion}}

\begin{figure} [!htb]
\begin {center}
\includegraphics[width=10cm] {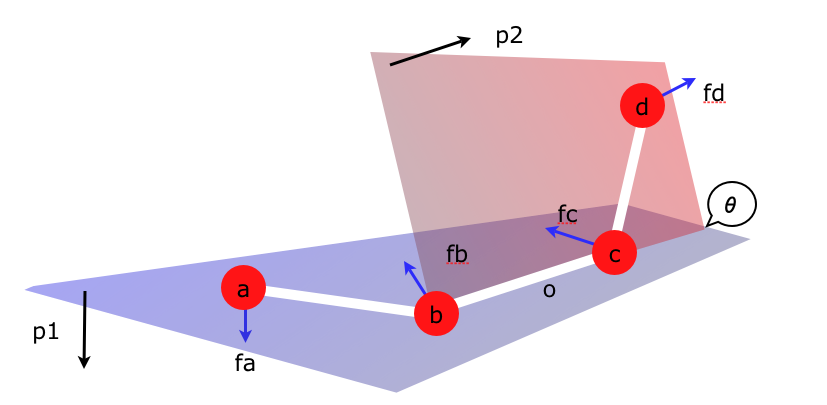}
\end {center}
\caption {Torsion angle $\theta$}
\label {figure:torsion}
\end{figure}

A torsion angle $\theta$ defined by four atoms $a,b,c,d$ is shown on
Fig. \ref {figure:torsion}.  In the OPLS \cite{DFTJ97} force-field, as in many other force-fields,
potentials of torsion angles have a {\it ``triple-cosine''} form. This means
that the potential $\cal U$ of a torsion angle $\theta$ is defined
by\footnote {
In the following, we will always omit the last parameter $A_4$.
}:
\begin {equation} \label{equation:dihe}
{\cal U}(\theta) = 0.5 [
   A_1 (1+cos (  \theta))
+ A_2 (1 - cos (2\theta)) 
+ A_3 (1 + cos (3\theta))
+ A_4
]
\end {equation}

The partial derivative of the torsion angle potential according to the
position $r_a$ of $a$ is:
\begin{equation}
{\frac {\partial {\cal U} (\theta)} {\partial r_a}} = 
{\frac {\partial {\cal U} (\theta)} {\partial \theta}} . 
{\frac {\partial \theta} {\partial r_a}} 
\end{equation}
The partial derivative of the potential according to the angle $\theta$ is:
\begin {eqnarray} \label{equation:dihf}
\frac {\partial {\cal U}(\theta)} {\partial \theta} 
&=& 0.5 (
     - A_1 sin (  \theta)
     + 2A_2 sin (2\theta) 
     - 3A_3 sin (3\theta)
)\\
&=& -0.5 (
     A_1 sin (  \theta)
     - 2A_2 sin (2\theta) 
     + 3A_3 sin (3\theta)
)
\end {eqnarray}

\subsection {Forces on a and d}
Let us call $\theta_1$ the angle $\widehat {abc}$. 
Atom $a$ turns around direction $bc$, on a circle of radius
$|ab|sin (\theta_1)$. The partial derivative of $\theta$ according to the
position of $a$ is:
\begin{equation}
{\frac {\partial \theta} {\partial r_a}} = \frac{1} {|ab|sin (\theta_1)}
\end {equation}

\noindent
We thus have:
\begin{equation}
{\frac {\partial {\cal U} (\theta)} {\partial r_a}} = 
\frac {-0.5} {|ab|sin (\theta_1)}
(
     A_1 sin (  \theta)
- 2A_2 sin (2\theta) 
+ 3A_3 sin (3\theta)) 
\end{equation}

\noindent
Similarly, for atom $d$, noting $\theta_2$ the angle
$\widehat {bcd}$~:
\begin{equation}
{\frac {\partial {\cal U} (\theta)} {\partial r_d}} = 
\frac {-0.5} {|cd|sin (\theta_2)}
(
     A_1 sin (  \theta)
- 2A_2 sin (2\theta) 
+ 3A_3 sin (3\theta))
\end{equation}

\noindent
Let $\mvector {p_1}$ the normalized vector orthogonal to the plane $abc$, and
$\mvector {p_2}$ the normalized vector orthogonal to the plane $bcd$
(the angle between $\mvector {p_1}$ and $\mvector {p_2}$  is $\theta$):
\begin{eqnarray}
\mvector {p_1} = \normal {\mvector {ba} \times \mvector {bc}}\\
\mvector {p_2} = \normal {\mvector {cd} \times \mvector {cb}}
\end{eqnarray}

\noindent
The force applied on $a$ is:
\begin{equation} \label {eq:torsion:fa}
\mvector {f_a} = \frac {0.5} {|ab|sin (\theta_1)} (
     A_1 sin (  \theta)
- 2A_2 sin (2\theta) 
+ 3A_3 sin (3\theta)).\mvector {p_1}
\end{equation}

\noindent
In the same way, the force applied on $d$ is:
\begin{equation} \label {eq:torsion:fd}
\mvector {f_d} = \frac {0.5} {|cd|sin (\theta_2)} (
     A_1 sin (  \theta)
- 2A_2 sin (2\theta) 
+ 3A_3 sin (3\theta)).\mvector {p_2}
\end{equation}

\subsection {Forces on b and c}
We now have to determine the forces $\mvector {f_b}$  and
$\mvector {f_c}$ to be applied on $b$ and $c$.  The equilibrium
conditions imply two constraints:
(A) the sum of the forces has to be null:
\begin{equation}\label{eq:dih:null}
\mvector {f_a} + \mvector {f_b} + \mvector {f_c} + \mvector {f_d} = 0
\end{equation}
and (B) the sum of torques also has to be
null\footnote {
It is not possible to simply define $\mvector {f_b} = -\mvector {f_a}$
and $\mvector {f_c}
= - \mvector {f_d}$, as the sum of torques would be non-null,
thus leading to an increase of potential energy.}.
Calling $o$ the center of bond $bc$, this means:
\begin{equation}\label {eq:dih:null-sum-torque}
\mvector {oa} \times \mvector {f_a} +
 \mvector {od} \times \mvector {f_d} +
 \mvector {ob} \times \mvector {f_b} +
 \mvector {oc} \times \mvector {f_c}  
= 0
\end{equation}

\noindent
From (\ref {eq:dih:null-sum-torque}) it results:
\begin{equation}
( \mvector {ob} + \mvector {ba} ) \times \mvector {f_a} +
( \mvector {oc}  + \mvector {cd} ) \times \mvector {f_d} +
 \mvector {ob} \times \mvector {f_b} +
 \mvector {oc} \times \mvector {f_c}  
= 0
\end{equation}

\noindent
and:
\begin{equation}
( -\mvector {oc} + \mvector {ba} ) \times \mvector {f_a}+
( \mvector {oc}  + \mvector {cd} ) \times \mvector {f_d} -
 \mvector {oc} \times \mvector {f_b} +
 \mvector {oc} \times \mvector {f_c}  
= 0
\end{equation}

\noindent
which implies:
\begin{equation}\label {eq:dih:sum1}
 \mvector {oc} \times ( -\mvector {f_a} + \mvector {f_d} - \mvector
{f_b} + \mvector {f_c}  )
+ \mvector {ba} \times \mvector {f_a} +  \mvector {cd}  \times
\mvector {f_d} 
= 0
\end{equation}

\noindent
From (\ref{eq:dih:null}) it results:
\begin{equation}\label {eq:dih:sum2}
-\mvector {f_a} + \mvector {f_d} - \mvector {f_b} + \mvector {f_c}  =
2 ( \mvector {f_d} + \mvector {f_c})
\end{equation}

\noindent
Substituting (\ref {eq:dih:sum2}) in (\ref {eq:dih:sum1}), one gets:
\begin{equation}
 \mvector {oc} \times ( 2 ( \mvector {f_d} + \mvector {f_c}) )
+ \mvector {ba} \times \mvector {f_a} +  \mvector {cd}  \times
\mvector {f_d} 
= 0
\end{equation}
thus:
\begin{equation}
 2\mvector {oc} \times \mvector {f_d} + 2 \mvector {oc} \times \mvector {f_c}
+ \mvector {ba} \times \mvector {f_a} +  \mvector {cd}  \times \mvector {f_d} 
= 0
\end{equation}
which implies:
\begin{equation}
 2 \mvector {oc} \times \mvector {f_c}  =
-2\mvector {oc} \times \mvector {f_d} -  \mvector {cd} \times \mvector {f_d} -  
    \mvector {ba} \times \mvector {f_a} 
\end{equation}
and we finally get the condition that the torque from $\mvector {f_c}$
should verify in order (\ref{eq:dih:null-sum-torque}) to be true:
\begin{equation} \label {eq:cpl}
\mvector{oc} \times \mvector{f_c} =
- (\mvector {oc} \times \mvector {f_d} + 0.5  \mvector {cd} \times \mvector {f_d} +
    0.5 \mvector {ba} \times \mvector {f_a})
\end{equation}
Let us state:
\begin{equation}
\mvector {t_c} = - (\mvector {oc} \times \mvector {f_d} + 0.5  \mvector {cd} \times \mvector {f_d} +
    0.5 \mvector {ba} \times \mvector {f_a})
\end{equation}

\noindent
Equation $\mvector {oc} \times \mvector {x}  = \mvector {t_c}$ has an infinity
of solutions in $\mvector x$, all having the same component perpendicular
to $\mvector {oc}$. We thus simply choose as solution the force
perpendicular to $\mvector
{oc}$ defined by: 
\begin{equation}\label {eq:torsion:fc} 
\mvector {f_c} = (1/ |oc|^2) \mvector {t_c} \times \mvector
{oc} 
\end{equation}

\noindent
Equation (\ref {eq:cpl}) is verified because:
\begin{equation}
\mvector {oc} \times \mvector {f_c} =  (1/ |oc|^2) \mvector {oc} \times (\mvector {t_c} \times \mvector {oc})
\end{equation}
thus\footnote {
if $u \bot v$, then $u \times (v \times u) = |u|^2 v$.
}~:
\begin{equation} 
\mvector {oc} \times \mvector {f_c} = 
 (1/ |oc|^2) |oc|^2 \mvector {t_c} = \mvector {t_c}
\end{equation}

The value of $\mvector {f_b}$ is finally deduced from equation
(\ref{eq:dih:null}) stating the equilibrium of forces:
\begin{equation} \label {eq:torsion:fb}
\mvector {f_b}= - \mvector {f_a}  - \mvector {f_c} - \mvector {f_d}
\end{equation}

We have thus determined four forces $\mvector {f_a},\mvector {f_b},\mvector
{f_c},\mvector {f_d}$ whose sum is null (\ref {eq:dih:null}) and whose sum of torques is
also null (\ref {eq:dih:null-sum-torque}).

\section {Lennard-Jones Potentials \label {section:lj}}
A Lennard-Jones (LJ) potential ${\cal U} (r)$ between two atoms placed
at distance $r$ is defined by\footnote{
this is the 6-12 form; other forms of LJ potentials exist.
}:

\begin {equation} \label {equation:lj-1}
{\cal U}(r) = 4 \epsilon 
[
  {( \frac {\sigma} {r} )}^{12} -
  {( \frac {\sigma} {r} )}^{6} 
]
\end {equation}

\noindent
In this definition, parameter
$\sigma$ is the distance at which the potential is null, and parameter
$\epsilon$ is the minimum of the potential (corresponding to the
maximum of the attractive energy). 


Stating $A = \sigma^{12}$ and $B = \sigma^{6}$, Eq. (\ref
{equation:lj-1}) becomes:

\begin {equation} \label {equation:lj-2}
{\cal U}(r) = 
4 \epsilon 
(
  \frac {A} {{r}^{12}} -
  \frac {B} {{r}^{6}}
)
\end {equation}

\noindent
The partial derivative of $\cal U$ according to distance is thus:
\begin {eqnarray}
{\frac {\partial {\cal U} (r)} {\partial r}} &=&
4 \epsilon
(  -12  \frac {A} {{r}^{13}} +   
   6    \frac {B} {{r}^{7}}
) \\
&=& 24 \epsilon
(  - 2  \frac {A} {{r}^{13}} +  
        \frac {B} {{r}^{7}}
) \\
&=& \frac {24 \epsilon}  {r}
(  -2  \frac {A} {{r}^{12}} +   
         \frac {B} {{r}^{6}}
) \\
&=& -\frac{24 \epsilon}  {r}
(  2  {(\frac {\sigma} {r})}^{12} -   
       {(\frac {\sigma} {r})}^{6}
) 
\end{eqnarray}
Let $a$ and $b$ be two atoms. The force on $a$ is:
\begin{equation} \label {eq:lj:fa}
\mvector {f_a} = \frac {24 \epsilon}  {r}
(  2  {(\frac {\sigma} {r})}^{12} -   
       {(\frac {\sigma} {r})}^{6}
).\mvector {u} 
\end{equation} 
where $\mvector {u}$  is the normalization of $\mvector {ba}$. 
From the action/reaction principle, one deduces that the force on $b$
should be the opposite of the force on $a$:
\begin{equation} \label {eq:lj:fb}
\mvector {f_b} = -\mvector {f_a}
\end{equation} 

According to the definition of $\mvector {f_a}$ and $\mvector {f_b}$,
the sum of the forces applied to $a$ and $b$ is null:
\begin{equation} \label {eq:lj:null-sum} 
\mvector {f_a} + \mvector {f_b} = 0
\end{equation} 
Note that, as for bonds, no torque is produced because the two forces are colinear.

\section {Resume \label {section:resume}}
The forces defined in the previous sections are summed up in the following table:

\begin {center}
\begin {tabular} {r|c|l} 
Bond $ab$      & \ref {eq:bond:fa}     &  $\mvector {f_a} =
-2k(r-r_0) .  {\mvector {u}}$  \\
                          & \ref {eq:bond:fb}      &  $\mvector {f_b} =  -\mvector {f_a}$   \\ \hline
Valence $abc$   & \ref {eq:valence:fa}  &  $\mvector {f_a} = -2k (\theta-\theta_0)/|ab|.\mvector {p_a}$    \\
                          &  \ref {eq:valence:fb} & $\mvector {f_b} =  - \mvector {f_a} - \mvector {f_c}$ \\ 
                          &  \ref {eq:valence:fc} & $\mvector {f_c} = -2k (\theta-\theta_0)/|bc|.\mvector {p_c}$ \\\hline
Torsion $abcd$ & \ref {eq:torsion:fa} & $\mvector {f_a} = \frac {0.5} {|ab|sin (\theta_1)} (
     A_1 sin (  \theta)
- 2A_2 sin (2\theta) 
+ 3A_3 sin (3\theta)).\mvector {p_1}$ \\
                        & \ref {eq:torsion:fb}& $\mvector {f_b}= - \mvector {f_a}  - \mvector {f_c} - \mvector {f_d}$ \\ 
                        & \ref {eq:torsion:fc}& $\mvector {f_c} =
                        (1/|oc|^2) . \mvector {t_c} \times \mvector {oc}$\\
                         & \ref {eq:torsion:fd}& $\mvector {f_d} = \frac {0.5} {|cd|sin (\theta_2)} (
     A_1 sin (  \theta)
- 2A_2 sin (2\theta) 
+ 3A_3 sin (3\theta)).\mvector {p_2}$ \\ \hline
LJ $ab$             & \ref {eq:lj:fa} & $\mvector {f_a} = \frac {24 \epsilon}  {r}
(  2  {(\frac {\sigma} {r})}^{12} -   
       {(\frac {\sigma} {r})}^{6}
).\mvector {u}$\\
                         & \ref {eq:lj:fb} & $\mvector {f_b} =
                         -\mvector {f_a}$ \\
\end {tabular}
\end {center}

\begin {itemize}

\item[{\bf Bond}] In Eq. \ref {eq:bond:fa}, $k$ is the bond strength
  constant, $r$ is the distance between atoms $a$ and $b$, and $r_0$
  is the equilibrium distance, for which energy is null.  Vector
  $\mvector u$ is defined by $\mvector u = \normal {\mvector {ba}}$.

\item[{\bf Valence}] In \ref {eq:valence:fa} and \ref {eq:valence:fc},
  $k$ is the angle strength constant, $\theta$ is the angle $\widehat
  {abc}$, and $\theta_0$ is the equilibrium angle, for which energy is
  null.
In \ref {eq:valence:fa}, 
  $\mvector {p_a}$ is defined by $\mvector {p_a} = \normal {\mvector
    {ba} \times (\mvector {ba} \times \mvector {bc})}$. In \ref
  {eq:valence:fc}, $\mvector {p_c}$ is defined by
  $\mvector {p_c} = \normal {\mvector {cb} \times (\mvector {ba}
    \times \mvector {bc})}$.

\item[{\bf Torsion}] In \ref {eq:torsion:fa} and \ref {eq:torsion:fd},
  $\theta$ is the torsion angle, $\theta_1$  is the angle $\widehat {abc}$,
  $\theta_2$  is the angle $\widehat {bcd}$ and $A_1$, $A_2$ and $A_3$
  are the
  parameters which define the ``three-cosine'' form of the torsion angle. Vector
  $\mvector {p_1}$ is defined by $\mvector {p_1} = \normal {\mvector {ba} \times
    \mvector {bc}}$ and $\mvector {p_2} = \normal {\mvector {cd} \times
    \mvector {cb}}$. In \ref {eq:torsion:fc}, $o$ is the middle of
  $bc$ and $\mvector {t_c}$ is defined by
 $\mvector {t_c}=
- (\mvector {oc} \times \mvector {f_d} + 0.5  \mvector {cd} \times \mvector {f_d} +
    0.5 \mvector {ba} \times \mvector {f_a})$.

  \item[{\bf LJ}] In \ref {eq:lj:fa}, $\sigma$  is the distance at which the 
    potential is null and $\epsilon$ is the depth of the potential
    (minimum of energy). As for bonds, one has
   $\mvector u = \normal {\mvector {ba}}$.

\end {itemize}

In each case (bond, valence, torsion, LJ interaction), the sum of the
forces that are applied to atoms is always null (Eq. (\ref
{eq:bond:null-sum}), (\ref {eq:valence:null-sum}),
(\ref{eq:dih:null}), (\ref {eq:lj:null-sum})).  Moreover, no torque is
induced by application of these forces: no torque is produced by bonds
and LJ interactions, as the produced forces are colinear; we have
verified in Sec. \ref {subsection:toques} that no torque is produced
by valence angles (\ref{eq:valence:torque}); for torsion angles, we
have chosen the forces in such a way that the sum of the forces and
the global sum of torques are always null (\ref
{eq:dih:null-sum-torque}).  This means that no energy is ever added
by the application of the forces during the simulation process.

\section {Conclusion \label {section:conclusion}}

We have precisely defined the forces that apply on atoms in MD simulations. The
definitions are given in a purely vectorial formalism (no use of a specific
coordinate system). We have shown that the sum of the forces and
the sum of the torques are always null, which means that
 the energy of molecular systems is preserved while the forces are applied.


\appendix

\section {Notations \label {section:notations}}

\begin {itemize}

\item if $a$ and $b$ are two atoms, we note $\mvector {ab}$ the vector
  with origin $a$ and end $b$; the distance between
  the two atoms is noted $|ab|$.

\item The null vector is noted $0$.

\item The length of vector $\mvector u$ is noted $| \mvector u
  |$. One thus has: $| \mvector {ab}  | = |ab|$.

\item Multiplication of $\mvector u$  by the scalar $n$ is noted $n
  . \mvector {u}$, or more simply $n \mvector {u}$.

\item The vectorial product of $\mvector u$ and $\mvector v$ is noted $\mvector {u} \times \mvector {v}$.

\item The scalar product of $\mvector u$ and $\mvector v$ is noted $\mvector {u} \bullet \mvector {v}$.

\item We write $\mvector u \bot \mvector v$ when $\mvector {u}$ and
  $\mvector {v}$ are orthogonal ($\mvector {u} \bullet \mvector {v} = 0$).

\item We note $\normal {\mvector {u}}$ the normalized vector from
  $\mvector u$  (same direction, but length equal to 1) defined by
$\normal {\mvector {u}} = (1/| \mvector u |).\mvector u$.

\item If $a$, $b$ and $c$ are atoms, we note $\widehat {abc}$ the
  angle formed by $a$, $b$ and $c$.

\end {itemize}


\end{document}